\documentstyle[12pt]{article}

\parskip=\medskipamount            
\def\7#1#2{\mathop{\null#2}\limits^{#1}}        

\def\beee{\begin{equation}}
\def\eeee{\end{equation}}
\def\dggg{^{\dagger}}
\begin{document}

\bibliographystyle{unsrt}


\begin{center}

{\bf SMALL VIOLATIONS OF STATISTICS}\\[.5in]

O.W. Greenberg\footnote{email address, owgreen@physics.umd.edu.}\\

Center for Theoretical Physics\\
Department of Physics \\
University of Maryland\\
College Park, MD~~20742-4111\\

\vspace{.5in}

University of Maryland Physics Paper 99-101\\
quant-ph/yymmddd
\end{center}

\vspace{5mm}

\begin{center}

Talk given at Orbis Scientiae 1998 to be published by Plenum Press.

\end{center}

\vspace{5mm}

\begin{abstract}
There are two motivations to consider statistics that are neither Bose
nor Fermi: (1) to extend the framework of quantum theory and of quantum
field theory, and (2) to provide a quantitative measure of possible 
violations of statistics.  After reviewing tests of statistics for 
various particles, and types of statistics that are neither Bose nor
Fermi, I discuss quons, particles characterized by the 
parameter $q$, which permit a
smooth interpolation between Bose and Fermi statistics;
$q=1$ gives bosons, $q=-1$ gives fermions.
The new result of this talk is work by Robert C. Hilborn and myself that
gives a heuristic argument for an extension of conservation
of statistics to quons with
trilinear couplings of the form $\bar{f}fb$, where
$f$ is fermion-like and $b$ is boson-like.  We showed that $q_f^2=q_b$.  
In particular, we related the bound
on $q_{\gamma}$ for photons to the bound on $q_e$ for electrons, allowing the very
precise bound for electrons to be carried over to photons.  An extension of 
our argument suggests
that all particles are fermions or bosons to high precision.
\end{abstract}

\begin{flushleft}
{\bf 1 INTRODUCTION}
\end{flushleft}
Michael Berry \cite{berry} reported on a very interesting new idea to derive the
connection of spin and statistics without using relativity in this session.  
After hearing
about this work it is going from the sublime to the ridiculous to consider
theories in which particles can have statistics that are neither Bose nor
Fermi.  Nonetheless, I will do so for two reasons: to stretch the framework
of quantum mechanics and of quantum field theory and to provide a formalism
that allows a quantitative measure of the accuracy with which a given particle
obeys either Bose or Fermi statistics.  For an earlier general discussion of violations
of statistics see \cite{ggg}.

I first review experiments that test statistics, and then survey the theoretical
ways in which violations of statistics can be introduced for identical particles.
I discuss quons, a type of particle that can have statistics that interpolate
continuously between bosons and fermions, in some detail \cite{q}.  At present, the 
quon theory
is the only theory that allows parametrization of small violations of statistics.
The new result that I report in this talk is conservation of statistics for quons
that relates the $q$-parameters for particles that couple to each other \cite{gh1}.  
For electrons and photons the result is $q_{photon}=q_{electron}^2$, which allows the
high-precision bound on possible violations of Fermi statistics for electrons to
be carried over to a comparably high-precision bound on violations of Bose statistics
for photons.  In conclusion, I mention the need for a refined derivation of the above
result.  I also state a result for the statistics of composite systems of quons that
Robert C. Hilborn and I found after the Orbis \cite{gh2}.

\begin{flushleft}
{\bf 2 EXPERIMENTS}
\end{flushleft}	
Until recently there were no high-precision tests of the Pauli exclusion principle
for fermions nor were there such tests for violations of Bose statistics
for bosons.  The exclusion principle is deeply engrained in our understanding
of quantum mechanics and there was no stimulus from either experiment or
theory to question it.  In the last few years, in part because of the great
success of the standard model, long-accepted features of the standard model,
such as Lorentz invariance \cite{li} and $CPT$ symmetry \cite{cpt}
have been questioned and, despite the absence of
experimental signals of violations, theories
have been advanced that allow violations or, if no violations are
seen, provide high-precision bounds on such violations for each type of particle.  
I am going to do the same for violations of statistics.

There are several types of experiments to detect violations of
Fermi or Bose statistics if they occur.  Here are three types: (i) search for
transitions among anomalous states--in either solids or in gases, (ii) search
for accumulation of particles in anomalous states, and (iii) search for 
deviations from the usual statistical properties of bulk matter.  R. Amado and H. 
Primakoff \cite{ap} pointed out that there is a superselection rule separating
states of identical particles in inequivalent irreducible representations of
the symmetric group, and because of this there are no transitions between
normal and anomalous states.  One has to look for transitions among anomalous
states rather than for transitions between
normal and anomalous states.  If transitions occur between states of the same
symmetry type, they occur with the normal rate.  Thus, for example, if the electrons
in an atom are not in a totally antisymmetric representation so that the $K$-shell
of the atom could have three electrons, then an electron in a higher shell would
make the transition to the $K$-shell at the usual electromagnetic rate. 

Atomic spectroscopy is the first place to search for violations of the 
exclusion principle since that is where
Pauli discovered it \cite{p}.  One looks for funny lines which
do not correspond to lines in the normal theory of atomic spectra.  There are
such lines, for example in the solar spectrum; however they probably can be
accounted for in terms of highly ionized atoms in an environment of high
pressure, high density and large magnetic fields.  Laboratory spectra are well
accounted for by theory and can bound the violation of the exclusion principle
for electrons by something like $10^{-6}$ to $10^{-8}$ using the parametrization I 
describe in the next paragraph.  

A useful
quantitative measure of the violation, ${v}$, is that ${v}$ is the 
coefficient of
the anomalous component of the two-particle density matrix; for fermions, 
the two-electron density matrix, $\rho_2$, is
\beee
\rho_2=(1-{v_F}) \rho_a+{v_F} \rho_s,               \label{rho2a}
\eeee
where $\rho_{a(s)}$ is the antisymmetric (symmetric) two-fermion density matrix.
Mohapatra and I surveyed a
variety of searches for violations of particle statistics in \cite{gm}.

Next I discuss an insightful experiment by Maurice and Trudy 
Goldhaber \cite{gold} that was designed to answer the question, ``Are the
electrons emitted in nuclear 
$\beta$-decay quantum mechanically identical to the
electrons in atoms?''  We know that the $\beta$-decay electrons have the same
spin, charge and mass as electrons in atoms; however the Goldhabers realized
that if the $\beta$-decay electrons were not 
quantum mechanically identical to those
in atoms, then the $\beta$-decay 
electrons would not see the $K$-shell of a heavy
atom as filled and would fall into the $K$-shell and emit an $x$-ray.  They
looked for such $x$-rays by letting $\beta$-decay electrons from a
natural source fall on a block of lead.  No such $x$-rays were found.  The 
Goldhabers were able to confirm that electrons from the two sources are indeed
quantum mechanically identical.  At the same time, they found that any
violation of the exclusion principle for electrons must be less than $5\%$.

E. Ramberg and G. Snow \cite{rs} developed this experiment into one which yields a
high-precision bound on violations of the exclusion principle.  Their idea was
to replace the natural $\beta$ source, which provides relatively few electrons,
by an electric current, in which case Avogadro's number is on our side.  The
possible violation of the exclusion principle is that a given collection of
electrons can, with different probabilities, be in different permutation
symmetry states.  The probability to be in the ``normal'' totally antisymmetric
state presumably would be close to one, 
the next largest probability would occur for the
state with its Young tableau having one row with two boxes, etc.  The idea of
the experiment is that each collection of electrons has a possibility of being
in an anomalous permutation state.  If the density matrix for a conduction 
electron together with the electrons in an atom has a projection onto such an
anomalous state, then the conduction electron will not see the $K$-shell of
that atom as filled. Then a transition into the $K$-shell with $x$-ray emission
is allowed.  Each conduction electron which comes sufficiently close to a given
atom has an independent chance to make such an $x$-ray-emitting transition, and
thus the probability of seeing such an $x$-ray is proportional to the number of
conduction electrons which traverse the sample and the number of atoms which the
electrons visit, as well as the probability that a collection of electrons can
be in the anomalous state.  Ramberg and Snow chose to run 30 amperes
through a thin copper strip for about a month.  They estimated the energy of the
$x$-rays which would be emitted due to the transition to the $K$-shell.  No 
excess of $x$-rays above background was found in this energy region.  Ramberg
and Snow set the limit 
\beee
v_F \leq 1.7 \times 10^{-26}.                  \label{vf}
\eeee
This is high precision, indeed!
K. Deilamian, J.D. Gillaspy and D.E. Kelleher \cite{dgh}
searched for transitions atoms of helium in which the two electrons
are in a symmetric state under permutations.  They used precision calculations
of the levels of such atoms made by G.W.F. Drake \cite{drake}.  They found the 
limit $v_F \leq 2 \times 10^{-7}$.  M. De Angelis, et al \cite{deangelis} and,
independently, R.C. Hilborn and C.L. Yuca \cite{hy} searched for forbidden bands
in the $O_2$ spectrum and found the bounds $v_B \leq 5 \times 10^{-7}$ and
$v_B \leq 5 \times 10^{-7}$, respectively, on violations of Bose statistics for
the oxygen nuclei.
Modugno, Ingusicio, and Tino \cite{modugno} 
found that the probability of finding the two $^{16}O$ nuclei
(spin 0) in carbon
dioxide in a permutation antisymmetric state is less than $5 \times 10^{-9}$.
Preliminary results on an experiment to bound violations of
Bose statistics for photons give $v_B \leq 10 \times 10^{-7}$ \cite{dd}.

\begin{flushleft}
{\bf 3. WAYS TO VIOLATE STATISTICS}
\end{flushleft}
It is difficult to violate the statistics of identical particles.  The
Hamiltonian must be totally symmetric in the dynamical variables of the 
identical particles; H cannot change the permutation symmetry type of the
wave function.  In particular, one cannot dial in a small violating term
using $H=H_S+\epsilon H_V$, since then the Hamiltonian would not be totally
symmetric.  Also one cannot, for example, have red electrons and blue electrons
even if there were only red electrons in our neighborhood.
This would lead to a doubling of the cross section 
{$\sigma(\gamma X \rightarrow e^+e^-X)$}, since photons couple universally.
\begin{flushleft}
{\bf 3.1 Gentile's Intermediate Statistics}
\end{flushleft}
The first attempt to go beyond Bose and Fermi statistics seems to have been
made by G. Gentile \cite{gentile}  who suggested an 
``intermediate statistics'' in which at
most $n$ identical particles could occupy a given quantum state.  In
intermediate 
statistics, Fermi statistics is recovered for $n=1$ and Bose statistics
is recovered for $n\rightarrow \infty$; thus intermediate statistics 
interpolates between Fermi and Bose statistics.  However Gentile's
statistics is not a proper quantum statistics because the condition of having
at most $n$ particles in a given quantum state is not invariant under change
of basis.  For example for intermediate statistics with $n=2$ the state
$|\psi \rangle=|k,k,k \rangle$ does not exist; however the state $|\chi
\rangle= 
\sum_{l_1,l_2,l_3}U_{k,l_1}U_{k,l_2}U_{k,l_3}|l_1,l_2,l_3 \rangle$ obtained
from $|\psi \rangle$ by the unitary change of single-particle basis 
$|k \rangle ^{\prime}=\sum_{l}U_{k,l}|l \rangle$ does exist.

By contrast, parafermi statistics of order $n$ (to be discussed just below) is
invariant under change of basis.  Parafermi statistics of order 
$n$ not only
allows at most $n$ identical particles in the same state, but also allows
at most $n$ identical particles in a symmetric state.  In the example just
described, neither $|\psi \rangle$ nor $|\chi \rangle$ exist for parafermi
statistics of order two.
\begin{flushleft}
{\bf 3.2 Green's Parastatistics}
\end{flushleft}	
H.S. Green \cite{hsgreen} proposed the first proper quantum mechanical
generalization of Bose and Fermi statistics.  Green noticed that the commutator
of the number operator with the annihilation and creation operators is the same
for both bosons and fermions
\beee
[n_k, a\dggg_l]_-=\delta_{kl}a\dggg_l.      \label{numbercr}
\eeee
The number operator can be written
\beee
n_k=(1/2)[a\dggg_k, a_k]_{\pm}+ {\rm const},       \label{number}
\eeee
where the anticommutator (commutator) is for the Bose (Fermi) case.  If these
expressions are inserted in the number operator-creation operator commutation
relation, the resulting relation is {\it trilinear} 
in the annihilation and creation operators.  Polarizing the number operator to
get the transition operator $n_{kl}$ that annihilates a free particle in state
$k$ and creates one in state $l$ leads to Green's trilinear commutation relation
for his parabose and parafermi statistics,
\beee
[[a\dggg_k, a_l]_{\pm}, a\dggg_m]_-=2\delta_{lm}a\dggg_k.     \label{tri}
\eeee
Since these rules are trilinear, the usual vacuum condition,
\beee
a_k|0\rangle=0,                                              \label{vac}
\eeee
does not suffice to allow calculation of matrix elements of the $a$'s and
$a\dggg$'s; a condition on one-particle states must be added,
\beee
a_k a\dggg_l|0\rangle=\delta_{kl}|0\rangle.        \label{vac2}
\eeee

Green found an infinite set of solutions of his commutation rules, one for each 
integer, using an ansatz in terms of Bose and Fermi
operators.  Let
\beee
a_k\dggg=\sum_{p=1}^n b_k^{(\alpha) \dagger},~~a_k=\sum_{p=1}^n b_k^{(\alpha)},
                                          \label{ansatz}
\eeee
and let the $b_k^{(\alpha)}$ and $b_k^{(\beta) \dagger}$ 
be Bose (Fermi) operators
for $\alpha=\beta$ but anticommute (commute) for $\alpha \neq \beta$ for the 
``parabose'' (``parafermi'') cases.  This ansatz clearly satisfies Green's
relation.  The integer $p$ is the order of the parastatistics.  The physical
interpretation of $p$ is that for parabosons $p$ is the maximum number of
particles that can occupy an antisymmetric state, while for parafermions $p$
is the maximum number of particles that can occupy a symmetric state (in
particular, the maximum number that can occupy the same state).  The case $p=1$
corresponds to the usual Bose or Fermi statistics.
Later Messiah
and I \cite{owg-mes} proved that Green's ansatz gives all Fock-like solutions of
Green's commutation rules.  Local observables have a form analogous to the usual
ones; for example, the local current for a spin-1/2 theory is 
$j_{\mu}=(1/2)[\bar{\psi}(x), \psi(x)]_-$.  From Green's ansatz, it is clear
that the squares of all norms of states are positive, since sums of Bose or
Fermi operators give positive norms.  Thus parastatistics gives a set of
orthodox theories.  Parastatistics is one of the
possibilities found by Doplicher, Haag and Roberts \cite{dhr} in a general study of
particle statistics using algebraic field theory methods.
Haag's recent book \cite{haag} gives a good review of this work.

This is all well and good; however the violations of statistics provided by
parastatistics are gross.  Parafermi statistics of order two has up to two
particles in each quantum state.  High-precision experiments are not necessary
to rule this out for the all particles we think are fermions.

\begin{flushleft}
{\bf 3.3 The Ignatiev-Kuzmin Model and ``Parons''}
\end{flushleft}	
Interest in possible small violations of the exclusion principle was revived by
a paper of Ignatiev and Kuzmin \cite{ik} in 1987.  They constructed a model of
one oscillator with three possible states: a vacuum state, a one-particle
state and, with small probability, a two-particle state.  They gave trilinear 
commutation relations for their
oscillator.  Mohapatra and I showed that the Ignatiev-Kuzmin oscillator could be
represented by a modified form of the order-two Green ansatz \cite{gm2}.  We suspected that
a field theory generalization of this model having an infinite number of
oscillators
would not have local observables and set
about trying to prove this.  To our surprize, we found that we could construct
local observables and gave trilinear relations that guarantee the locality 
of the current \cite{gm2}.  We also checked the positivity of the norms with
states of three or fewer particles.  At this stage, we were carried away with
enthusiasm, named these particles ``parons'' since their algebra is a
deformation of the parastatistics algebra, and thought we had found a local
theory with small violation of the exclusion principle.
We did not know that Govorkov \cite{gov} had shown in
generality that any deformation of the Green commutation relations necessarily
has states with negative squared norms in the Fock-like representation.
For our model the first such negative-probability state occurs for
four particles in the representation of ${\cal S}_4$ with three boxes in the
first row and one in the second.  We were able to understand Govorkov's result
qualitatively as follows \cite{gm3}:
Since parastatistics of order $p$ is related by a
Klein transformation to a model with exact $SO(2)$ or $SU(2)$ internal symmetry,
a deformation of parastatistics that interpolates between Fermi and parafermi
statistics of order two would be equivalent to interpolating between the trivial
group whose only element is the identity and a theory with 
$SO(2)$ or $SU(2)$ internal symmetry.  This is impossible, since there is no
such interpolating group.
\begin{flushleft}
{\bf 3.4 Apparent Violations of Statistics Due to Compositeness}
\end{flushleft}	
Before getting to ``quons,'' the final type of statistics I will discuss, I want
to interpolate some comments about apparent violations of statistics due to
compositeness.  Consider two $^3 He$ nuclei, each of which is a fermion.  If
these two nuclei are brought in close proximity, the exclusion principle will
force each of them into excited states, plausibly with small amplitudes for the
excited states.  Let the creation operator for the nucleus at location $A$ be
\beee
b_A\dggg=\sqrt{1-\lambda_A^2}b_0\dggg+\lambda_A b_1\dggg+\cdots, |\lambda_A|<<1,
                                   \label{nuca}
\eeee
and the creation operator for the nucleus at location $B$ be 
\beee
b_B\dggg=\sqrt{1-\lambda_B^2}b_0\dggg+\lambda_B b_1\dggg+\cdots, |\lambda_B|<<1.
                                  \label{nucb}
\eeee
Since these nuclei are fermions, the creation operators obey fermi statistics,
\beee
[b_i\dggg, b_j\dggg]_+=0              \label{fermi}
\eeee
Then, 
\beee
b_A\dggg b_B\dggg|0\rangle
=[\sqrt{1-\lambda_A^2}\lambda_B-\lambda_A\sqrt{1-\lambda_B^2}]
b_0\dggg b_1\dggg |0\rangle,             \label{viol}
\eeee
\beee
\|b_A\dggg b_B\dggg|0\rangle \|^2 \approx (\lambda_A-\lambda_B)^2<<1,
                                      \label{norm}
\eeee
so with small probability, 
the two could even occupy the same location, because each could be excited
into higher states with different amplitudes.  This is not an intrinsic
violation of the exclusion principle but rather only an apparent violation due
to compositeness.

\begin{flushleft}
{\bf 4 QUONS}

{\bf 4.1 Quon Algebra and Fock Representation}
\end{flushleft}	
Now I come to my last topic, quons \cite{q}. The quon algebra is
\beee
a_k a_l \dggg-q a_l \dggg a_k=\delta_{kl}.            \label{quon}
\eeee
For the Fock-like representation I impose the vacuum condition
\beee
a_k |0\rangle=0.                                     \label{qvac}
\eeee

These two conditions determine all vacuum matrix elements of polynomials in the
creation and annihilation operators.  In the case of free quons all
non-vanishing vacuum matrix elements must have the same number of annihilators
and creators.  For such a matrix element with all annihilators to the left and
creators to the right, the matrix element is a sum of products of 
``contractions'' of the form $\langle 0|a a\dggg |0 \rangle$ just as in the case
of bosons and fermions.  The only difference is that the terms are multiplied by
integer powers of $q$.  The power can be given as a graphical rule:  Put
$\circ$'s
for each annihilator and $\times$'s for each creator in the order in which 
they occur in the matrix element on the $x$-axis.  
Draw lines above the $x$-axis connecting the pairs that are
contracted.  The minimum number of times these lines cross 
is the power of $q$ for that term in the matrix element.  Thus a modified
Wick's theorem holds for quon operators.

The physical significance of $q$ for small violations of Fermi statistics is
that $q=2 {v_F} -1$, where the parameter ${v_F}$ appears in Eq.(\ref{rho2a}).
For small violations of Bose statistics, the two-particle density matrix is
\beee
\rho_2=(1-{v_B}) \rho_s+{v_B} \rho_a,            \label{rho2s}
\eeee
where $\rho_{s(a)}$ is the symmetric (antisymmetric) two-boson density matrix.
Then $q=1- 2{v_B}$.

For $q$ in the open interval $(-1, 1)$ all
representations of the symmetric group occur.  As $q \rightarrow 1$ the
symmetric representations are more heavily weighted and at $q=1$ 
only the totally
symmetric representation remains; correspondingly, as $q \rightarrow -1$ the
antisymmetric representations are more heavily weighted and at $q=-1$ only the 
totally antisymmetric representation remains.  Thus for a general 
$n$-quon state there
are $n!$ linearly independent states for $-1<q<1$, but there is only one 
state for $q= \pm 1$.
I emphasize something that many people find very strange: {\it there is no
commutation relation between two creation or between two annihilation 
operators,} except for $q= \pm 1$, which, of course, correspond to Bose and 
Fermi statistics.  Indeed, the fact that the general $n$-particle state with
different quantum numbers for all the particles has $n!$ linearly independent
states proves that there is no such commutation relation between any number 
of creation
(or annihilation) operators. An even stronger statement holds:  There is no 
two-sided ideal containing a term with only creation operators.
Note that here quons differ from the ``quantum plane'' in which
\beee
xy=qyx                                      \label{qplane}
\eeee
holds.

Quons are an operator realization of the ``infinite statistics'' that were found as
a possible statistics by Doplicher, Haag and Roberts \cite{dhr} in their general
classification of particle statistics.  The simplest case, $q=0$ \cite{0},
suggested to me by Hegstrom \cite{heg}, 
was discussed earlier in the context of operator algebras by Cuntz \cite{cuntz}.
It seems
likely that the Fock-like representations of quons for $|q|<1$ are homotopic to
each other and, in particular, to the $q=0$ case, which is particularly simple. 
Thus it is convenient, as I will now do, 
to illustrate qualitative properties of quons for this simple case.  All
bilinear observables can be constructed from the number operator, 
$n_k \equiv n_{kk}$, or the
transition operator, $n_{kl}$, that obey 
\beee
[n_k, a\dggg_l]_-=\delta_{kl}a\dggg_l,
~~[n_{kl}, a\dggg_m]_-=\delta_{lm}a\dggg_k.            \label{qncr}
\eeee
Although the formulas for $n_k$ and $n_{kl}$ in the general case  are
complicated, the corresponding formulas for $q=0$ are simple \cite{0}.
Once Eq.(\ref{qncr}) holds, the Hamiltonian and other observables can be
constructed in the usual way; for example for free particles
\beee
H=\sum_k \epsilon_k n_k,~~ {\rm etc.}  \label{qh}
\eeee
The obvious thing is to try
\beee
n_k=a\dggg_k a_k.   \label{qntry}
\eeee
Then
\beee
[n_k,a\dggg_l]_-=\delta_{kl} a\dggg_k -a\dggg_la\dggg_ka_k.  \label{qntry2}
\eeee
The first term in Eq.(\ref{qntry2}) is $\delta_{kl}a\dggg_k$ as desired; however
the second term is extra and must be canceled.  This can be done by adding the
term $\sum_ta\dggg_ta\dggg_ka_ka_t$ to the term in Eq.(\ref{qntry}).  This cancels
the extra term, but adds a new extra term, that must be canceled by another
term.  This procedure yields an infinite series for the number operator
and for the transition operator,
\beee
n_{kl}=a\dggg_ka_l+\sum_ta\dggg_ta\dggg_ka_la_t+\sum_{t_1,t_2}a\dggg_{t_2}
a\dggg_{t_1}a\dggg_ka_la_{t_1}a_{t_2}+ \dots   \label{qninfty}
\eeee
As in the Bose case, this infinite series for the transition or number 
operator defines an unbounded operator whose domain includes states made by
polynomials in the creation operators acting on the vacuum.
(As far as I know, this is the first case in which the number operator, 
Hamiltonian, etc. for a free field are of infinite degree.  Presumably this is
due to the fact that quons are a deformation of an algebra and are related to
quantum groups.)
For nonrelativistic theories, the $x$-space form of the transition operator
is \cite{physica}
\begin{eqnarray}
\rho_1({\bf x};{\bf y})=\psi\dggg({\bf x})\psi({\bf y})    
+\int d^3z\psi\dggg
({\bf z})\psi\dggg({\bf x})\psi({\bf y})\psi({\bf z})  \nonumber \\
+\int d^3z_1d^3z_2\psi({\bf
z_2})\psi\dggg({\bf z_1})\psi\dggg({\bf x})\psi({\bf y})\psi({\bf z_1})
\psi({\bf z_2})+ \cdots,  \label{qtran}
\end{eqnarray} 
which obeys the nonrelativistic locality requirement
\beee
[\rho_1({\bf x};{\bf y}),\psi\dggg({\bf w})]_-=\delta({\bf y}-{\bf
w})\psi\dggg({\bf x}),~~ {\rm and}~~ 
\rho({\bf x};{\bf y})|0\rangle=0.  \label{qloc}
\eeee
The apparent nonlocality of this formula associated with the space integrals has
no physical significance.  To support this last statement, consider
\beee
[Qj_{\mu}(x),Qj_{\nu}(y)]_-=0,~~x \sim y,   \label{qcurrent}
\eeee
where $Q=\int d^3x j^0(x)$.  Equation (\ref{qcurrent}) seems to have nonlocality
because of the space integral in the $Q$ factors; however, if 
\beee
[j_{\mu}(x),j_{\nu}(y)]_-=0,~~x \sim y, \label{lcurrent}
\eeee
then Eq.({\ref{qcurrent}) holds, despite the apparent nonlocality.  What is relevant
is the commutation relation, not the representation in terms of a space
integral.  (The apparent nonlocality of quantum electrodynamics in the Coulomb
gauge is another such example.)

In a similar way,
\beee
[\rho_2({\bf x}, {\bf y};{\bf y}^{\prime}, {\bf x}^{\prime}), 
\psi\dggg({\bf z})]_-=\delta({\bf x}^{\prime}-{\bf z})
\psi\dggg({\bf x})\rho_1({\bf y},{\bf y}^{\prime})+
\delta({\bf y}^{\prime}-{\bf z})
\psi\dggg({\bf y})\rho_1({\bf x},{\bf x}^{\prime}).   \label{rho2cr}
\eeee
Then the Hamiltonian of a nonrelativistic theory with two-body interactions has
the form
\beee
H=(2m)^{-1} \int d^3x \nabla _x \cdot \nabla_{x^{\prime}} 
\rho_1({\bf x}, {\bf x}^{\prime})|_{{\bf x}={\bf x}^{\prime}} +
\frac{1}{2} \int d^3x d^3y V(|{\bf x}-{\bf y}|) 
\rho_2({\bf x},{\bf y};{\bf y}, {\bf x}).              \label{ham2}
\eeee
\begin{eqnarray}
[H,\psi\dggg({\bf z}_1) \dots \psi\dggg({\bf z}_n)]_-=
[-(2m)^{-1} \sum_{j=1}^n \nabla^2_{{\bf z}_i}+ 
 \sum _{i<j} 
V(|{\bf z}_i-{\bf z}_j|)] \psi\dggg({\bf z}_1) \dots \psi\dggg({\bf z}_n)
\nonumber \\
+\sum_{j=1}^n \int d^3x V(|{\bf x}-{\bf z}_j|)\psi\dggg({\bf z}_1) 
\cdots \psi\dggg({\bf z}_n)\rho_1({\bf x}, {\bf x}^{\prime}).   \label{ham2cr}
\end{eqnarray}
Since the last term on the right-hand-side of Eq.(\ref{ham2cr}) vanishes when the
equation is applied to the vacuum, this equation shows that the usual
Schr\"odinger equation holds for the $n$-particle system.  Thus the usual
quantum mechanics is valid, with the sole exception that any permutation
symmetry is allowed for the many-particle system.  This construction justifies
calculating the energy levels of (anomalous) atoms with electrons in states
that violate the exclusion principle using the normal Hamiltonian, 
but allowing anomalous
permutation symmetry for the electrons \cite{drake}.
\begin{flushleft}
{\bf 4.2 Positivity of Squares of Norms}
\end{flushleft}	
I have not yet addressed the question of positivity of the squares of norms
that caused grief in the paron model.  Several authors have given proofs of
positivity \cite{zag,boz-spe,speicher,fivel}.  The proof of Zagier provides an 
explicit formula for the determinant of the $n! \times n!$ matrix 
of scalar products among the states of $n$ particles in different quantum
states.  Since this determinant is one for $q=0$, the norms will be positive
unless the determinant has zeros on the real axis.  Zagier's formula
\beee
det\; M_{n}(q) =\Pi_{k=1}^{n-1}(1-q^{k(k+1)})^{(n-k)n!/k(k+1)},  \label{zag}
\eeee   
has zeros only on the unit circle, so the desired positivity follows.
Although quons satisfy the requirements of nonrelativistic locality, the quon
field does not obey the relativistic requirement, namely spacelike
commutativity of observables.  Since quons interpolate smoothly between
fermions, which must have odd half-integer spin, and bosons, which must have
integer spin, the spin-statistics theorem, which can be proved, at least for
free fields, from locality would be violated if locality were to hold for quon
fields.  It is amusing that, nonetheless, the free quon field obeys the TCP
theorem and Wick's theorem holds for quon fields \cite{q}.
\begin{flushleft}
{\bf 4.3 Speicher's ansatz}
\end{flushleft}	
Speicher \cite{speicher} has given an ansatz for the Fock-like representation of quons 
analogous to Green's ansatz for parastatistics.  Speicher represents the quon
annihilation operator as
\beee
a_k={\rm lim}_{N \rightarrow \infty}N^{-1/2}\sum_{\alpha=1}^N b_k^{(\alpha)},    \label{weak}
\eeee
where the $b_k^{(\alpha)}$ are Bose oscillators for each $\alpha$, but with
relative commutation relations given by
\beee
b_k^{(\alpha)} b_l^{(\beta) \dagger}=s^{(\alpha, \beta)}b_l^{(\beta) \dagger}
b_k^{(\alpha)}, \alpha \neq \beta,~ {\rm where}~~ s^{(\alpha, \beta)}= \pm 1.  \label{weakcr}
\eeee
Equation(\ref{weak}) is taken as the weak limit, $N \rightarrow \infty$, 
in the vacuum expectation state of the 
Fock space representation of the $b_k^{(\alpha)}$.  
In this respect, Speicher's ansatz differs from Green's, which is an operator
identity.  Further to get the Fock-like representation of the quon algebra, 
Speicher chooses a probabilistic condition for the signs 
$s^{(\alpha, \beta)}$,
\beee
{\rm prob}(s^{(\alpha, \beta)}=1)=(1+q)/2,     \label{probs}
\eeee
\beee
{\rm prob}(s^{(\alpha, \beta)}=-1)=(1-q)/2.     \label{proba}
\eeee
Since a sum of Bose operators acting on
a Fock vacuum always gives a positive-definite norm, the positivity property is
obvious with Speicher's construction.

Speicher's ansatz leads to the conjecture that there is an 
infinite-valued hidden degree of freedom underlying $q$-deformations analogous
to the hidden degree of freedom underlying parastatistics.

If one asks ``How well do we know that a given particle obeys Bose or Fermi statistics?," 
we need a quantitative way to answer the question.  That requires a formulation in which either
Bose or Fermi statistics is violated by a small amount.  As stated earlier, we 
cannot just add to the Hamiltonian a small term 
that violates Bose or Fermi statistics; such a term would not be invariant
under permutations of the identical particles and thus would clash with the particles being
identical.  As mentioned above parastatistics, which does violate Bose or Fermi statistics,
gives gross violations.  The only way presently available to allow small violations
of statistics is the quon theory just described.

Unfortunately, the quon theory is not completely satisfactory.
The observables in quon theory do not commute at spacelike
separation.  If they did, particle statistics could change continuously from Bose to Fermi
without changing the spin.  Since spacelike commutativity of observables leads to the
spin-statistics theorem, this would be a direct contradiction.  Kinematic Lorentz 
invariance can be maintained, but without spacelike commutativity or anticommutativity
of the fields the theory may not be consistent.  

For nonrelativistic theories, however, quons are consistent.  The nonrelativistic
version of locality is
\begin{equation}
[\rho({\bf x}),\psi({\bf y})]=-\delta({\bf x}-{\bf y})\psi({\bf y})               \label{nrcr}
\end{equation}
for an observable $\rho({\bf x})$ and a field $\psi({\bf y})$ 
and this does hold for quon theories.  It is the antiparticles that prevent locality in
relativistic quon theories.

\begin{flushleft}
{\bf 5. CONSERVATION OF STATISTICS}
\end{flushleft}	
\begin{flushleft}
{\bf 5.1 Conservation of Statistics for Bosons and Fermions}
\end{flushleft}	
The first conservation of statistics theorem states that terms in the Hamiltonian density 
must have an even number of Fermi fields and that composites of fermions and bosons are 
bosons, unless they contain an odd number of fermions, in which case they are 
fermions 
\cite{wig,ehr}.
\begin{flushleft}
{\bf 5.2 Conservation of Statistics for Parabosons and Parafermions}
\end{flushleft}	
The extension to parabosons and parafermions is more complicated \cite{owg-mes}; 
however, the main
constraint is that for each order $p$ at least two para particles must enter into every
reaction. 

Reference \cite{qg} argues that the condition that the energy of widely
separated subsystems be additive requires that all terms in the
Hamiltonian be ``effective Bose operators'' in that sense that
\begin{equation}
[{\cal H}({\bf x}), \phi({\bf y})]_-\rightarrow 0, |{\bf x}-{\bf y}| \rightarrow \infty.   
                                             \label{ascr}
\end{equation}
For example,  ${\cal H}$ should not have a term such as 
$\phi(x){\psi(x)}$, where ${\phi}$ is 
Bose and $\psi$ is Fermi, because then the contributions to the 
energy of widely separated subsystems would alternate in sign.  Such terms are
also prohibited by rotational symmetry.  This discussion was given in the context of
external sources.  

It is well known that external fermionic sources must be multiplied by a
Grassmann number in order to be a valid term in a Hamiltonian.  This is
necessary, because additivity of the energy of widely separated systems requires
that all terms in the Hamiltonian must be effective Bose operators.  I
constructed the quon analog of Grassmann numbers \cite{qg} in order to allow
external quon sources.  Because this issue was overlooked, the bound on
violations of Bose statistics for photons claimed 
in \cite{fiv-ext} is invalid. 

For a fully quantized field theory, one can replace Eq.(\ref{ascr})
by the asymptotic causality condition, asymptotic local commutativity,
\begin{equation}
[{\cal H}({\bf x}), {\cal H}({\bf y})]_-=0, |{\bf x}-{\bf y}| \rightarrow \infty \label{decr}
\end{equation}
or by the stronger causality condition, local commutativity,
\begin{equation}
[{\cal H}({\bf x}), {\cal H}({\bf y})]_-=0, {\bf x} \neq {\bf y}.           \label{decr2} 
\end{equation}
Studying this condition for quons in electrodynamics is complicated, 
since the terms in the interaction
density will be cubic.  It is simpler to use the description of the electron current or
transition operator as an external source represented by a quonic Grassmann number.
\begin{flushleft}
{\bf 5.3 Conservation of Statistics for Quons}
\end{flushleft}	
Here we 
give a heuristic argument for conservation of statistics for quons based on a simpler
requirement in the context of quonic Grassmann external sources \cite{gh1}.
The commutation relation of the quonic photon operator is          
\begin{equation}
a(k) a^{\dagger}(l) -q_{\gamma} a^{\dagger}(l)  a(k)=\delta(k-l),   \label{qphoton}
\end{equation}
where $q_{\gamma}$ is the $q$-parameter for the photon quon field.
We call the 
quonic Grassmann numbers for the electron transitions to which the photon quon operators
couple $c(k)$.
The Grassmann numbers that serve as the external source for coupling
to the quon field for the photon must obey
\begin{equation}
c(k) c(l)^{\star}  -q_{\gamma} c(l)^{\star}  c(k)=0,       \label{qgrass}
\end{equation}
and the relative commutation relations must be
\begin{equation}
a(k) c(l)^{\star} -q_{\gamma} c(l)^{\star}  a(k)=0,        \label{relative}
\end{equation}
etc.  Since the electron current
for emission or absorption of a photon with transition of the electron from one atomic
state to another is bilinear in the creation and annihilation operators for the electron,
a more detailed description of the photon emission would treat the photon as
coupled to the electron current, rather than to an external source.  
We impose the requirement that the leading terms in the commutation relation for the
quonic Grassmann numbers of the source that couples to the photon should be mimicked by 
terms bilinear in the electron operators. The electron operators obey the relation
\begin{equation}
b(k) b^{\dagger}(l)- q_{e}b^{\dagger}(l) b(k)=\delta(k-l),     \label{qe}
\end{equation}
where $q_e$ is the $q$-parameter for the electron quon field.  

To find the connection between $q_e$ and $q_{\gamma}$ we make the following associations,
\begin{equation}
c(k) \Rightarrow b^{\dagger}(p)b(k+p), ~~ c^{\star}(l) \Rightarrow b^{\dagger}(l+r)b(r)
                                                                   \label{replace}
\end{equation}
We now replace the $c$'s in Eq.(\ref{qgrass}) with the products of operators given in 
Eq.(\ref{replace}) and obtain
\begin{equation}
[b^{\dagger}(p) b(k+p)] [b^{\dagger}(l+r) b(r)]
- q_{\gamma} [b^{\dagger}(l+r) b(r)][b^{\dagger}(p) b(k+p)]=0.      \label{newcr1}                                  
\end{equation}
This means that the source $c(k)$
is replaced by a product of $b$'s that destroys net momentum $k$; the source $c^{\star}(l)$
is replaced by a product of $b$'s that creates net momentum $l$.  
We want to rearrange the operators in the first term of Eq.(\ref{newcr1}) to match the second
term, because this corresponds to the standard normal ordering for the transition operators.
For the products $b b^{\dagger}$ we use Eq.(\ref{qe}).  For the products $bb$, as mentioned
above, there is no operator relation; however {\it on states in the Fock-like representation}
there is an approximate relation,
\begin{equation}
b(k+p) b(r) = q_e b(r) b(k+p) + {\rm ~terms ~of ~order} ~~1-q_e^2.         \label{approx}
\end{equation}
In other words, in the limit $q_e \rightarrow -1$, we retrieve the usual anticommutators        
for the electron operators.  (The analogous relation for an operator that is approximately
bosonic would be that the operators commute in the limit $q_{bosonic} \rightarrow 1$.)
We also use the adjoint relation
\begin{equation}
b^{\dagger}(p) b^{\dagger}(l+r)=q_eb^{\dagger}(l+r) b^{\dagger}(p) + 
{\rm terms ~of ~order} ~~1-q_e^2      \label{adj}
\end{equation}
and, finally,
\begin{equation} 
q_e b^{\dagger}(p) b(r)=b(r) b^{\dagger}(p) - \delta(r-p).     \label{adja}
\end{equation}
We require only that the quartic terms that correspond to the quonic Grassmann relation
Eq.(\ref{qgrass}) cancel, so we drop terms in which either $k+p=l+r$ or $r=p$. We also drop
terms of order $1-q_e^2$.  In this approximation, we 
find that Eq.(\ref{newcr1}) becomes
\begin{equation}
(q_e^2-q_{\gamma})[b^{\dagger}(l+r) b(r)][b^{\dagger}(p) b(k+p)] \approx 0,  \label{newapprox}                             \label{newcr}
\end{equation}
and conclude that 
\begin{equation}
q_e^2 \approx q_{\gamma}.           \label{result}
\end{equation}
This relates the bound on violations of Fermi statistics for electrons to the bound on violations
of Bose statistics for photons and allows the extremely precise bound on
possible violations of Fermi statistics for 
electrons to be carried over to photons.  Eq.(\ref{result})
is the quon analog of the conservation of statistics relation that the square of the
phase for transposition of a pair of fermions equals the phase for 
transposition of a pair of bosons.

Arguments analogous to those just given, based on the source-quonic photon relation, 
Eq.(\ref{relative}), lead to 
\begin{equation}
q_{e\gamma}^2 \approx q_{\gamma},      \label{relativeq}
\end{equation}
where $q_{e\gamma}$ occurs in the relative commutation relation
\begin{equation}
a(k) b^{\dagger}(l)=q_{e\gamma} b^{\dagger}(l) a(k).      \label{abrel}
\end{equation}
Since the normal commutation relation between Bose and Fermi fields is 
for them to commute \cite{araki},
this shows that $q_{e\gamma}$ is close to one.

\begin{flushleft}
{\bf 6. HIGH-PRECISION BOUNDS}
\end{flushleft}	
Since the Ramberg-Snow bound on Fermi statistics for electrons is 
\begin{equation}
v_e \leq 1.7 \times 10^{-26}  \Longleftrightarrow  q_e \leq -1 + 3.4 \times 10^{-26},
                                   \label{rsbound}
\end{equation}  
the bound on Bose statistics for photons is
\begin{equation}
q_{\gamma} \geq 1 - 6.8 \times 10^{-26}  \Longleftrightarrow 
v_{\gamma} \leq 3.4 \times 10^{-26}.
                                      \label{pbound}
\end{equation} 
This
bound for photons is much stronger than could be gotten by a direct experiment.
Nonetheless
D. DeMille and N. Derr are performing an experiment that promises to give the best
{\it direct} bound on Bose statistics for photons \cite{dd}.
It is essential to test every basic property in as direct a way as possible.  Thus 
experiments that yield direct
bounds on photon statistics, such as the one being carried out by 
DeMille and Derr, are important.

Teplitz, Mohapatra and Baron have suggested a method to set a very low limit on
violation of the Pauli exclusion principle for neutrons \cite{tmb}.

The argument just given that the $q_e$ value for electrons implies $q_{\gamma}\approx
q_e^2$ for photons
can be run in the opposite direction to find 
$q{_\phi}^2 \approx q_{\gamma}$ for each charged field
$\phi$ that couples bilinearly to photons.  Isospin and other symmetry arguments then imply 
that
almost all particles obey Bose or Fermi statistics to a precision comparable to the 
precision with which electrons obey Fermi statistics.

\begin{flushleft}
{\bf 7 CONCLUSION}
\end{flushleft}	
In concluding, we note that further work should be carried out to justify the approximations
made in deriving Eq.(\ref{result}) and also
to derive the relations among the $q$-parameters that follow from couplings that do not 
have the form ${\bar f}fb$.  We plan to return to this topic in a later paper.
After the Orbis, Hilborn and I derived a generalization of the 
Wigner--Ehrenfest-Oppenheimer rule of the statistics of bound states in terms
of the quon statistics of their constituents, 
$q_{composite}=q_{constituent}^{n^2}$, where $n$ is the number of constituents
in the bound state \cite{gh2}.

\begin{flushleft}
{\bf Acknowledgements}
\end{flushleft}	

I thank the Aspen Center for Physics for a visit during which part of
this work was carried out.  The unique atmosphere of the Center encourages concentrated
work without the distractions of one's home university. This work was supported in part 
by the National Science Foundation.  The work on conservation of statistics for
quons was done in collaboration with Robert C. Hilborn.\\
\vspace{.6cm}

\vglue .2cm

\end{document}